\documentclass[twocolumn,pra,aps,showpacs,superscriptaddress,amssymb]{revtex4}
\usepackage{graphicx}

\begin{document}

\title{Laser cooling of a trapped two-component Fermi gas}

\author{Z. Idziaszek}
\affiliation{Institut f\"ur Theoretische Physik, Universit\"at Hannover, 
D-30167 Hannover,Germany}
\affiliation{Centrum Fizyki Teoretycznej, Polska Akademia Nauk, 02-668 Warsaw, Poland}
\author{L. Santos}
\affiliation{Institut f\"ur Theoretische Physik, Universit\"at Hannover, D-30167 Hannover,Germany}
\author{M. Baranov}
\affiliation{Institut f\"ur Theoretische Physik, Universit\"at Hannover, D-30167 Hannover,Germany}
\affiliation{Russian Research Center Kurchatov Institute, Kurchatov Square, 123182 Moscow, Russia}
\author{M. Lewenstein}
\affiliation{Institut f\"ur Theoretische Physik, Universit\"at Hannover, D-30167 Hannover,Germany} 

\begin{abstract}
We study the collective Raman cooling of a trapped two-component Fermi gas using  
quantum master equation in the {\it festina lente} regime, 
where the heating due to photon reabsorption can be neglected. 
The Monte Carlo simulations show, that 3D temperatures of the order of 
$0.008$~$T_F$ can be achieved. We analyze the heating related to background losses, and show
that our laser-cooling scheme can maintain the temperature of the gas without significant additional losses. 
\end{abstract}


\maketitle

The realization of Bose-Einstein condensation (BEC) \cite{BEC} and degenerated Fermi gases 
\cite{Fermions} has outbursted the interest in the physics of 
ultracold bosonic and fermionic gases. A Fermi gas with  attractive interactions
undergoes  for temperatures $T$ below a critical one, $T_c$, a transition into the superfluid
Bardeen-Cooper-Schrieffer (BCS) phase \cite{DeGennes}. 
The accomplishment of BCS and its possible detection have been 
considered in detail  (see e.g. Refs. \cite{BCS}). 
Unfortunately, evaporative  cooling currently employed in experiments
is based on collisional processes, and
has not yet allowed to reach  $T<T_c$, since $T_c$ is much 
smaller than the Fermi temperature $T_F$, and for $T<T_F$ the collisions
 are strongly suppressed due to Pauli blocking \cite{HollandF}. 
Recently, however, it has been proposed that $T_c$ may be significantly 
increased by employing Feshbach resonances \cite{Holland}, or optical lattices \cite{OptLatt}. 
This opens the possibility to achieve BCS 
at a temperature where the collisions are still efficient enough. 
\\ \indent
In this Rapid Communication 
we study the laser cooling of a two-component Fermi gas in the  
{\it festina lente} (FL) regime \cite{Festina}, where the spontaneous emission rate $\gamma$ is smaller 
than the trap frequency $\omega$. In this regime the heating due to photon 
reabsorption is prevented, as recently observed for tightly bound atoms in optical lattices \cite{Weiss}.
The laser cooling in the FL regime has been already predicted to work for bosons \cite{CoolBosons} 
and for polarized fermions \cite{fermicool}. By using Monte Carlo simulations, we show here 
that laser cooling allows for bringing a two component Fermi system to $T\ll T_F$.
\\ \indent
The laser cooling of fermions toward $T\ll T_F$ is obstacled by
several problems. One of them is the inhibition of spontaneous emission, which
results in the decrease of the cooling efficiency. In Ref.
\cite{fermicool} we have predicted that the inhibition problem
can be overcome by either dynamically adjusting the spontaneous emission
rate in a Raman cooling process, or by employing specially designed
anharmonic traps. In this paper we apply the former solution. Another obstacle is related to
the inelastic losses that create holes deeply in the Fermi sea, producing a significant heating. 
We show that background collisions do
not affect significantly the laser cooling. In fact, our cooling method  
can be employed to maintain a two-component Fermi gas at a fixed $T$, 
for a relatively long time in a trap. Finally, we derive an analytic formula that describes the 
heating of the trapped gas due to background collisions. The heating is shown to be smaller than in the 
homogeneous case.
\\ \indent
We consider fermionic atoms with an accessible electronic three-level $\Lambda$ scheme, containing
states $|g\rangle$, $|e\rangle$ and $|r\rangle$.
The ground state $|g\rangle$ is coupled via a Raman transition 
to the metastable state $|e\rangle$. The latter state is also coupled by an optical transition
to the upper state $|r\rangle$, from which atoms rapidly decay into $|g\rangle$. 
The adiabatic elimination of $|r\rangle$, leads to an effective two-level system, characterized by 
tunable parameters: Rabi frequency $\Omega$, and spontaneous emission rate $\gamma$. The 
value of $\gamma$ can be controlled by modifying the coupling from $|e\rangle$ to $|r\rangle$. 

The atoms are placed in a dipole trap characterized by a Lamb-Dicke parameter 
$\eta=2\pi a/\lambda$, with $a=\sqrt{\hbar/2m\omega}$ being the size of the ground state of the trap, and 
$\lambda$ the laser wavelength. A dipole trap was recently used for the all-optical
production of a degenerate gas of two Li species \cite{Granade}, and in current experiments 
in Mg \cite{Ernst}. We consider a spherically symmetric trap 
with incomensurable frequencies $\omega^g$ and $\omega^e$, for the ground and the excited 
state respectively. The latter assumption simplifies the dynamics of the 
spontaneous emission processes in the FL limit. The cooling process consists of 
sequences of Raman pulses of frequencies, adjusted in such a way that they 
induce the transition of atoms to the lower motional states of the trap.
We assume that the power of cooling lasers is sufficiently weak, and therefore during each pulse no 
significant population in $|e\rangle$ is present. This allows to eliminate 
adiabatically the level $|e\rangle$, and to consider only the density matrix
$\rho(t)$ for the atoms in $|g\rangle$, and 
being diagonal in the Fock representation corresponding to the bare trap levels.

We assume that the laser acts only on one component, whereas  
the other one is cooled sympathetically, which is sufficient to reach $T\ll T_F$. 
Using the standard theory of quantum-stochastic processes \cite{GardinerB,Carmichael} we derive  
the quantum master equation (ME) for the density matrix $\rho(t)$ in a similar way as that for bosons \cite{CoolBosons}:
\begin{equation}
\dot\rho(t)={\cal L}_0\rho+{\cal L}_1\rho+{\cal L}_2\rho,
\label{ME}
\end{equation}
where ${\cal L}_0\rho=-i\hat H_{\mathrm{eff}}\rho(t)+i\rho(t)\hat H_{\mathrm{eff}}^{\dag}+{\cal J}\rho(t)$, 
${\cal L}_1\rho=-i[\hat H_{\mathrm{las}},\rho(t)]$, 
${\cal L}_2\rho=-i[\hat H_{\mathrm{coll}},\rho(t)]$.
The first term in the ME (\ref{ME}) describes the evolution of the atoms in the vacuum of the
electromagnetic field. The effective non-Hermitian Hamiltonian $H_{\mathrm{eff}}$  is defined as 
\begin{eqnarray}
\hat H_{\mathrm{eff}}&=&\sum_{\bf m}\omega_{\bf m}^{g} g_{\bf m}^{\dag}g_{\bf m}+\sum_{\bf l}(\omega_{\bf l}^{e}
-\delta)e_{\bf l}^{\dag} e_{\bf l} + {} \nonumber \\
& & {} + \sum_{\bf n} \omega_{\bf n}^{b} b_{\bf n}^{\dag} b_{\bf n}
- i \frac{\gamma}{2} \sum_{{\bf l},\bf m} \xi_{\,\bf lm} e_{\bf l}^{\dag} g_{\bf m} g_{\bf m}^{\dag} 
e_{\bf l},
\label{Heff}
\end{eqnarray}
and the so-called jump super-operator ${\cal J}$ is given by
\begin{equation}
{\cal J}\rho(t)= \sum_{{\bf l},\bf m} \xi_{\,\bf lm} g_{\bf m}^{\dag}e_{\bf l}\rho(t)e_{\bf l}^{\dag}
g_{\bf m}.
\label{Jump}
\end{equation}
Here $g_{\bf m}$ ($e_{\bf l}$) is the annihilation operator of atoms of the first component, in the 
ground (excited) state, and in the trap level ${\bf m}$ (${\bf l}$). The annihilation operator of atoms of
the second component, and in the trap level ${\bf n}$, is denoted by $b_{\bf n}$. These operators 
fulfill the standard fermionic anticommutation relations: 
$\{g_{\bf m},g_{\bf n}^{\dag}\}=\{e_{\bf m},e_{\bf n}^{\dag}\}=\{b_{\bf m},b_{\bf n}^{\dag}\}=
\delta_{{\bf m},\bf n}$. 
In Eq. (\ref{Heff}) $\gamma$ denotes the single-atom effective spontaneous emission rate, 
$\omega _{\bf m}^g$, $\omega_{\bf m}^e$, $\omega_{\bf m}^b$ are the energies of state ${\bf m}$ of 
the trap:
$|g\rangle$, $|e\rangle$, $|b\rangle$ respectively, and $\delta$ is the laser detuning from the atomic 
transition. The coefficients $\xi_{\,\bf lm}$ are defined as follows
$\xi_{\,\bf lm}= \int_{0}^{2\pi} d\phi \int_{0}^{\pi} d\theta \, \sin\theta \, {\cal W}(\theta,\phi) 
|\eta_{\,\bf lm}(\vec k)|^{2}$, 
where ${\cal W}(\theta,\phi)$ is the fluorescence dipole pattern, and 
$\eta_{\,lm}(\vec{k})=\langle e,{\bf l}|e^{i \vec{k}\cdot \vec{r}}|g,{\bf m}\rangle$
the Frank-Condon factors.
 
The interaction of the laser light with the atoms is governed by the Hamiltonian $\hat H_{\mathrm{las}}$:
\begin{equation}
\hat H_{\mathrm{las}} = \frac{\Omega}{2}\sum_{{\bf l},\bf m}\eta_{\,\bf lm}(\vec{k}_{L})e_{\bf l}^{\dag}g_{\bf m}
+H.c., 
\label{Hlas}
\end{equation}
where $\Omega$ is the Rabi frequency and $\vec{k}_L$ is the wavevector of the cooling laser. 
Binary collisions are described by 
\begin{equation}
\hat H_{\mathrm{coll}} = \sum_{{\bf m},{\bf n},{\bf q},{\bf p}} 
U_{{\bf m},{\bf n},{\bf q},{\bf p}} \, g_{\bf m}^{\dag} g_{\bf n} b_{\bf q}^{\dag} b_{\bf p}, \label{Hcoll}
\end{equation}
where due to Fermi statistics only collisions between different species are allowed.
We neglect the collisions with the atoms in $|e\rangle$, since 
only a small fraction of atoms is excited in each pulse.
The collisional amplitudes are defined as 
$U_{{\bf m},{\bf n},{\bf q},{\bf p}}=(4\pi\hbar a_{\mathrm{sc}}/m) \int_{R^3}d^3x \,
\phi_{\bf m}^{\ast}({\bf x}) \phi_{\bf n}({\bf x}) \beta_{\bf q}^{\ast}({\bf x}) \beta_{\bf p}({\bf x})$, 
where $\phi_{\bf n}$ ($\beta_{\bf n}$) denotes the wavefunction of the state ${\bf n}$ in the $|g\rangle$, 
($|b\rangle$) trap, and $a_{\mathrm{sc}}$ is the scattering length.

We require the FL condition
$\gamma < \omega$ at least in the initial phase of the cooling process, and also assume that
$\Omega<\gamma$ and $\Omega^2/\gamma \ll \omega$, which allows for the adiabatic elimination of 
$|e\rangle$, performed by means of projection operator techniques \cite{GardinerB}. 
For weak interactions between the two species, the dynamics due to the collisions and 
the dynamics due to the laser-cooling are independent \cite{Coll}. The laser cooling 
may be described by the ME without the collisional part, whereas the collisional part takes
the form of a quantum Boltzmann master equation (QBME) \cite{QBME}. 
The probability of a laser-induced transition from state ${\bf m}$ to state ${\bf n}$ of the trap $|g\rangle$
is given by 
\begin{eqnarray}
P_{\bf m\rightarrow n}^{\mathrm{opt}} &=& \frac{\Omega^{2}}{\gamma} N_{\bf m}^{1} (1-N_{\bf n}^{1}) 
\times {} \nonumber \\
& & {} \times \sum_l \frac{ \gamma^2 \xi_{\bf ln} 
|\eta_{\,\bf lm}({\vec k}_L)|^2} {4 [\delta-(\omega_{\bf l}^e-\omega_{\bf m}^g)]^2+\gamma^2 
{R_{\bf ml}}^2},
\label{Pmn}
\end{eqnarray}
where $R_{\bf ml} =  \sum_{\bf n} \xi_{\,\bf ln} (1-N_{\bf n}^{1}+\delta_{{\bf n},\bf m})$ are the factors modifying 
the spontaneous emission rate in a Fermi gas, 
and $N_{\bf m}^{1}$ and $N_{\bf m}^{2}$ denotes the number of atoms occupying the state ${\bf m}$ of the trap 
$|g\rangle$ and $|b\rangle$, respectively.
In the regime of quantum degeneracy, coefficients $R_{\bf ml}$ vanish, inhibiting the spontaneous emission 
and forcing the atoms to remain excited for a long time. 
This prolongs the cooling process, and can result in excited-ground collisions, 
which lead to heating and losses. In addition, the adiabatic elimination used in the derivation of 
Eq. (\ref{Pmn}) ceases to be valid. The cooling efficiency is also decreased due to 
the fermionic inhibition factor $(1-N_{\bf n})$ in the numerator of (\ref{Pmn}).
The negative influence of the statistics can be overcome by dynamically increasing $\gamma$ 
during the Raman cooling \cite{fermicool}, in order to avoid 
the inhibition effects, but still remaining in the FL regime: $\gamma R_{\bf ml}< \omega$. 
Still, a small fraction of the atoms will remain in the excited state after the cooling pulses, 
and has to be removed from the trap in order to avoid 
inelastic collisions. The latter aim can be achieved by optically pumping the 
excited atoms to a third non-trapped level. 

The probability of a collision between two fermions of different species, from 
the states ${\bf n}$ and ${\bf p}$ to the states 
${\bf m}$ and ${\bf q}$, respectively, is given by   
\begin{eqnarray} 
P_{{\bf n},{\bf p} \rightarrow {\bf m},{\bf q}}^{\mathrm{coll}} & = & \frac{\pi}{\omega} 
N_{\bf n}^{1} N_{\bf p}^{2} (1-N_{\bf m}^{1}) (1-N_{\bf q}^{2}) \times {} \nonumber \\
& & {} \times \left|U_{{\bf m},{\bf n},{\bf q},{\bf p}}\right|^2
\delta_{E_{\bf n}+E_{\bf p},E_{\bf m}+E_{\bf q}}.
\label{Pcoll}
\end{eqnarray}
The fermionic inhibition factors $(1-N_{\bf m}^{1})$ also slow the collisional processes.

In our simulations we assumed the atoms as confined in an isotropic, 3D  
harmonic trap with frequency $\omega=\omega^{g}=\omega^{e}=\omega^{b}$. Due to the limitations 
when simulating relatively large systems ($N \sim 10^4$), 
we employ ergodic approximation, i.e. we assume that the populations of the states with the same 
energy are equal. This approximation 
relies on the fact that thermalization inside the same energy shell is much faster 
than between different energy shells. In addition we assume
that the collisional processes is much faster than the laser cooling. The detailed calculation of  
the transition probabilities within ergodic approximation is presented in Ref. \cite{fermicool2}.
Due to numerical limitations we assumed $\eta = 2$, which could be e.g. the case of 
potassium atoms in a dipole trap with $\omega = 2 \pi \times 2.4$~kHz 
(employed in Li experiments \cite{Granade}), and a laser wavelength 
$\lambda \simeq 720$~nm. We consider a scattering length for the interactions
between the two species $a_{\mathrm{sc}}=157 a_0$, where $a_0$ is the Bohr radius. This
value corresponds to the interactions between $|F=9/2,m_F=9/2\rangle$ and
$|F=9/2,m_F=7/2\rangle$ of $^{40}$K \cite{DeMarco}. The trap is assumed to have 
the same depth for both species, and contains $81$ energy levels ($91881$ states).
We assume both components to have equal initial number of atoms, $N=10660$,  
corresponding to a Fermi energy $E_F = 38 \hbar \omega$. For this relatively large number of atoms,
the dynamics generated by the collisional part of the ME equation leads to an equilibrium distribution, which
agrees very well with the one calculated from the grand canonical ensemble \cite{fermicool2}. 
Hence we start the simulations from a thermal distribution. 
%%%%%%%%%%%%%%
\begin{figure}
\begin{center}
\includegraphics[width=6.8cm,clip]{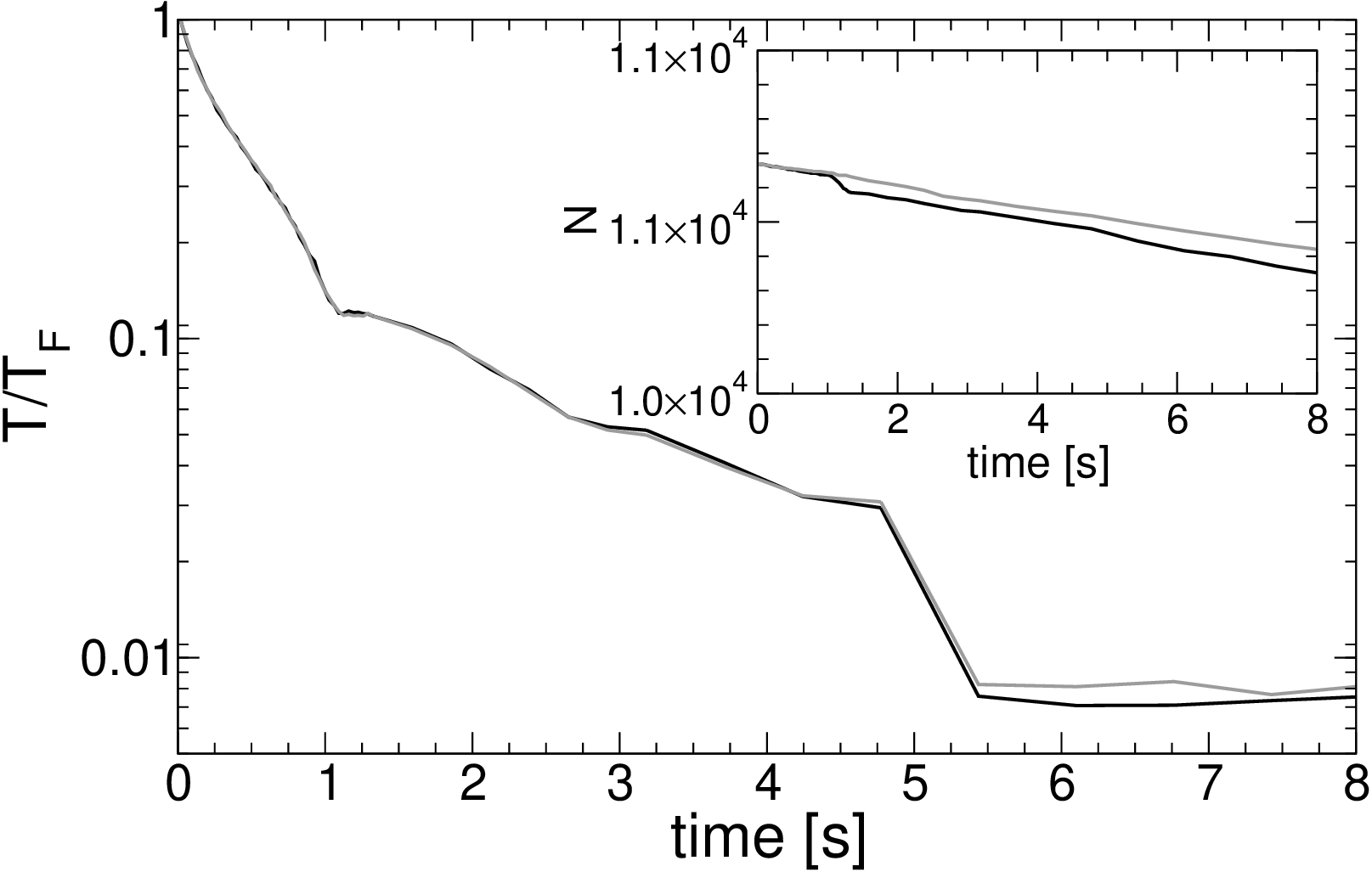}
\end{center}
\vspace{-0.5cm}
\caption{
\label{fig:1}
Time-dependence of the temperatures of the two components (darker and lighter curves) during the cooling
by controlling the effective spontaneous emission rate. 
Inset: time dependence of the number of atoms. The losses are due to background collisions and 
the removal of long-living excited atoms. For details, see text.}
\end{figure}
%%%%%%%%%%%%%%
%%%%%%%%%%%%%
\begin{figure}
\begin{center}
\includegraphics[width=6.8cm,clip]{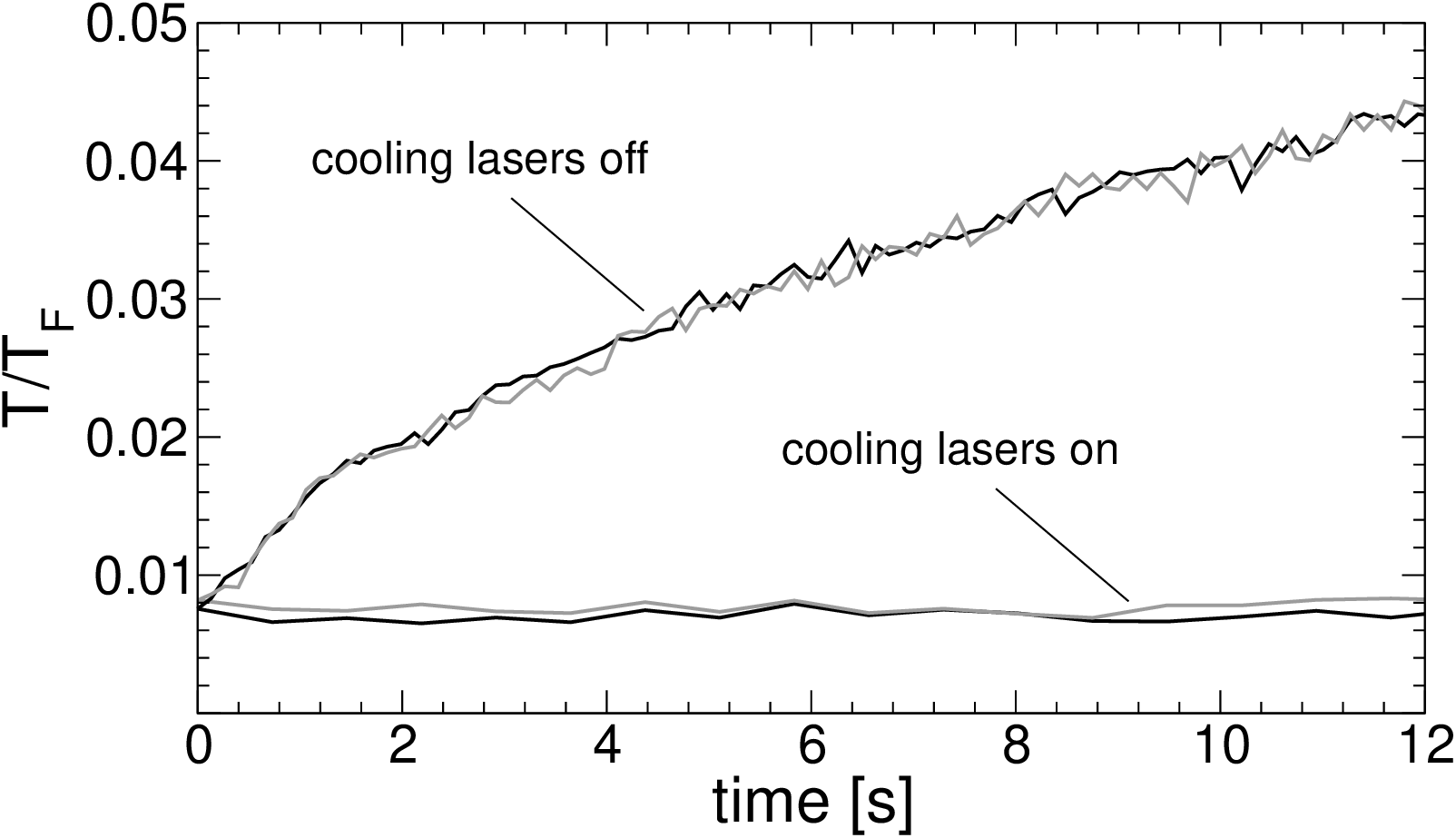}
\end{center}
\vspace{-0.5cm}
\caption{
\label{fig:2}
Evolution of a degenerate two-component Fermi gas after the cooling process, 
in the presence of background collisions with a rate $\gamma_{\mathrm{bg}}=350$~Hz.
The temperatures of the two components are plotted (darker and lighter curves).}
\end{figure}
%%%%%%%%%%%%%%%%
%%%%%%%%%
\begin{figure}[hb]
\begin{center}
\includegraphics[width=6.8cm,clip]{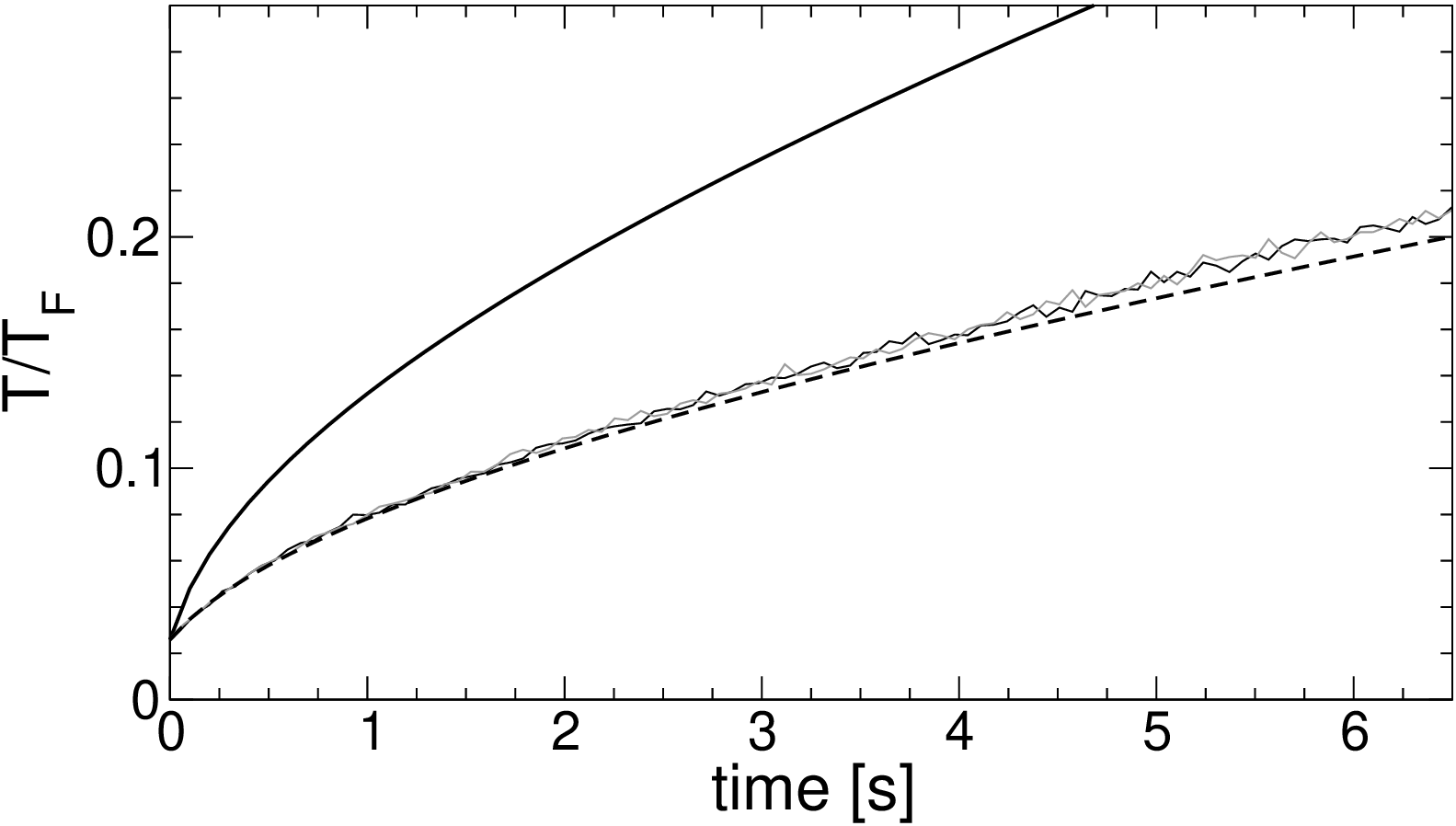}
\end{center}
\vspace{-0.5cm}
\caption{
\label{fig:3}
Heating of the degenerate two-component Fermi gas due to background collisions with a rate
$\gamma_{\mathrm{bg}}=10$~Hz.
The plot shows the temperature of the two components
 obtained in the simulation (darker and lighter thin solid curves), the analytic result 
of Ref. \cite{Timmer} calculated for a homogeneous system (thick solid line), and the data calculated 
from Equation (\ref{Temp}) (thick dashed line).}
\end{figure}
%%%%%%%%%%%%%

In a Fermi gas the main loss sources are provided by background collisions, resulting from non ideal 
vacuum conditions, and photoassociation. The photoassociation losses 
(when the laser is tuned between molecular resonances) are typically of the order of $10^{-14}$cm$^3$/s for 
laser intensities of $1$mW/cm$^2$ \cite{Machholm}. In our case the applied laser intensities 
are typically $1000$ times smaller and we estimate that for $N=10660$ atoms, the atomic density is
smaller than $3.5 \times 10^{14}$cm$^{-3}$. Therefore the photoassociation losses can be safely neglected.
On the contrary, the background collision apart from decreasing the number of atoms, 
generate holes deep within the Fermi sea. Those holes, after subsequent thermalization,
may lead to a significant heating \cite{Timmer}.
We have assumed that background losses depopulate each state of the trap with the rate 
$\gamma_{\mathrm{bg}}$, which is independent of the energy of the state 
\begin{equation}
\dot{N}_n^j=-\gamma_{\mathrm{bg}} N_n^j,
\label{Nbg}
\end{equation}
where $j=1,2$ enumerates the components. In the simulations we assume 
$\gamma_{\mathrm{bg}} = 1/350$ Hz \cite{HollandF}. We include also in the simulations the losses produced by 
the removal of long-living excited atoms at the end of each cooling pulse.

Fig.~\ref{fig:1} shows the evolution of the temperature of the laser-cooled two-component Fermi gas. 
Initially we consider $T_0=T_F$.
The cooling process was divided into four stages, each one consisting of a sequence of two Raman pulses.  
The employed pulses are characterized by the following parameters: detuning   
$\delta/\omega=\{(-11,-12)$, $(-16,-17)$, $(-19,-20)$,$(-25,-26)\}$ respectively, Rabi frequency
$\Omega/\gamma=\{(0.113,0.113)$, $(0.008,0.012)$, $(0.0025,0.004)$,$(0.003,0.01)\}$ respectively, and length
$\Delta t/\omega^{-1}=\{(250,250)$, $(2000,2000)$, $(4000,4000)$,$(4000,6500)\}$ respectively. 
For these parameters not more than $10\%$ of the atoms is excited during each pulse, and thus 
the conditions of the adiabatic elimination are fulfilled. The temperature was determined by fitting
the calculated distribution of fermions to a thermal distribution. As one can observe,
a final temperature $T\simeq 0.008 T_F$ may be reached within $8$~s. The losses associated with the 
cooling process do not exceed $2$\% and are slightly larger for the laser-cooled component, due to 
the removal of the long-living excited atoms (inset of Fig.~\ref{fig:1}). 

We have analyzed the effect of the losses on the laser-cooled gas. To this aim we have considered an initial 
gas at $T=0.008 T_F$ (end of the cooling process in Fig.~\ref{fig:1}), and compared two different cases: 
(i) the laser is turned off, and the gas is heated due to the background collisions, (ii) the laser is 
turned on, and the cooling pulses of the last stage are continuously applied. 
Fig.~\ref{fig:2} presents the evolution of the temperature of both species for these two cases.
As observed, the laser cooling compensates for the heating induced by the creation of 
holes in the degenerate distribution. Hence, it helps to maintain the degenerate gas for a relatively 
long time in the trap. We have verified that the continuous application of the laser cooling does not lead 
to substantially larger losses.

Finally, we have analyzed the results for the heating due to background collisions 
with the help of an analytical model. For the background losses that decrease the population
of the trap levels in the way described by Eq. (\ref{Nbg}), one can calculate the evolution of the 
temperature analytically \cite{Timmer}. To this end, it is sufficient to observe that the 
Eq. (\ref{Nbg}) do not change the mean energy per particle $\varepsilon$, which has to
remain constant in the absence of cooling: $\varepsilon\big(N(t),T(t)\big)=$~{\it const}.
At low temperatures $\tilde{T}=T/T_F\ll 1$, the mean energy per particle in a harmonic trap is given by
$\varepsilon(N,T) = (3/4) E_F(N) + (\pi^2/2) (k_B T)^2/E_F(N)$.
Using $\varepsilon\big(N(t),T(t)\big)=\varepsilon(N_0,T_0)$, 
where $N_0$ is the initial number of atoms, and $T_0$ is the initial temperature of the gas, 
we obtain 
\begin{equation}
\tilde{T}(t)=\sqrt{\frac{E_F(N_0)}{E_F(N(t))} \tilde{T}_0^2+\frac{3}{2 \pi^2}\left(
\frac{E_F(N_0)}{E_F(N(t))}-1\right)}.
\label{Temp}
\end{equation}
For a large system $E_F(N)\simeq \hbar \omega (6 N)^{1/3}$, and 
$N(t) \simeq N_0 \exp(-\gamma_{\mathrm{bg}}t)$, hence
\begin{equation}
\tilde{T}(t)=\sqrt{e^{\gamma_{\mathrm{bg}}t/3} \tilde{T}_0^2+ 3 \left(
e^{\gamma_{\mathrm{bg}}t/3}-1\right)/2 \pi^2}.
\label{Temp1}
\end{equation}
Fig.~\ref{fig:3} presents the evolution of the temperature for a rate
$\gamma_{\mathrm{bg}}=10$~Hz, which is much larger than the 
typical experimental rates, and was employed to reduce the amount of time-consuming numerical 
calculations. Fig.~\ref{fig:3} compares our numerical results, the predictions of 
the analytical formula of \cite{Timmer} derived for a homogeneous system, and the data calculated from 
Eq. (\ref{Temp}) with $E_F(N)\simeq \hbar \omega \big((6 N)^{1/3}-2\big)$, 
which is more appropriate for a finite size system \cite{TimmHarm}.
%\footnote{Although Ref.~\cite{Timmer} contains also the formula derived
%for a harmonic trap, we have not included it in this comparison, since in our opinion 
%it contains an incorrect factor.}. 
The analytic curve fits 
very well to the numerical data. The small discrepancy for larger temperatures originate from the fact, 
that the expression for the mean energy, which we use,
is valid up to order $(T/T_F)^2$.
From the plot, it is clearly seen that the increase of the temperature in a trapped gas 
is much slower that in the homogeneous case.
\\ \indent
In conclusion, we have studied the laser cooling of trapped two-component Fermi gases. Our method 
exploits the dynamical adjusting of the effective spontaneous emission rate in Raman cooling. 
Using Monte Carlo simulations, we have shown 
that laser cooling is able to cool the two-component Fermi system below $0.01$~$T_F$.
We have also discussed the losses which may affect the laser-cooled gas.  
In this context, we have shown that our laser-cooling scheme 
can be employed to maintain a two-component Fermi gas at a fixed temperature in the presence of background collisions, 
without significant additional atom losses.
Finally we have derived an analytic formula for the temperature of a trapped Fermi gas heated by 
background collisions. We have shown that the heating rate is smaller than in the homogeneous case.

We acknowledge support from the Alexander von Humboldt Stiftung, 
the Deutsche Forschungsgemeinschaft, the RTN Cold Quantum gases, 
the ESF Program BEC2000+, the Polish KBN Grant No. 5-P03B-103-20,
and the Russian Foundation for Fundamental Research.


\begin{thebibliography}{99}
\bibitem{TimmHarm} Although Ref.~\cite{Timmer} contains also the formula derived
for a harmonic trap, we have not included it in this comparison, since in our opinion 
it contains an incorrect factor.
\end{thebibliography}
\end{document}